\documentclass[Journal]{IEEEtran}
\usepackage{graphicx}
\usepackage{amsmath, amssymb}
\usepackage{hyperref}
\usepackage[normalem]{ulem}
\useunder{\uline}{\ul}{}
\usepackage{float}  
\usepackage{wrapfig}
\usepackage{placeins}

\title{Can Machine Learning Algorithms Outperform Traditional Models for Option Pricing?}

\author{
	\IEEEauthorblockN{Georgy Milyushkov} \\
	\IEEEauthorblockA{École polytechnique, France \\ 
		\texttt{georgy.milyushkov@polytechnique.edu}}
}

\begin{document}
	\title{Can Machine Learning Algorithms Outperform Traditional Models for Option Pricing?}
	
	\author{
		\IEEEauthorblockN{Georgy Milyushkov} \\
		\IEEEauthorblockA{\textit{École polytechnique}, France \\
			\texttt{georgy.milyushkov@polytechnique.edu}}
	}
	
	\maketitle
	
	\begin{abstract}
		This study investigates the application of machine learning techniques, specifically Neural Networks, Random Forests, and CatBoost for option pricing, in comparison to traditional models such as Black-Scholes and Heston Model. Using both synthetically generated data and real market option data, each model is evaluated in predicting the option price. The results show that machine learning models can capture complex, non-linear relationships in option prices and, in several cases, outperform both Black-Scholes and Heston. These findings highlight the potential of data-driven methods to improve pricing accuracy and better reflect market dynamics.
	\end{abstract}
	

\section{Introduction}
Accurate option pricing plays a fundamental role in financial markets, enabling investors, traders, and institutions to assess risk, construct hedging strategies, and identify arbitrage opportunities. Classical pricing models, such as the Black-Scholes formula \cite{BSart}, offer closed-form solutions for European options under highly strict assumptions, including constant volatility, continuous trading, and frictionless markets. In practice, however, real market prices frequently deviate from these theoretical values due to stochastic volatility, market frictions, and non-linear dynamics.\\

A major limitation of the Black-Scholes framework is its assumption of constant volatility, which fails to capture well-documented empirical phenomena such as volatility smiles and clustering. More advanced models, such as the Heston model \cite{Heston}, introduce stochastic volatility in an effort to better reflect observed market behavior. While these models improve fit by incorporating richer dynamics, they remain parametric in nature and are constrained by their underlying assumptions.\\

Machine learning (ML) techniques offer a fundamentally different approach. Rather than implementing a strict functional form, ML models learn pricing relationships directly from data. This flexibility makes them particularly well-suited for capturing non-linear dependencies, structural breaks, and market inefficiencies that traditional models often miss. Recent advances in computational power and algorithm design have made it feasible to apply ML methods to financial problems at scale, prompting growing interest in their application to option pricing.\\

This study investigates the application of machine learning techniques, specifically: Neural Networks (MLP), Random Forest, and CatBoost for option pricing. The objective is to assess whether ML models can approximate the stochastic processes governing option prices more effectively than traditional methods, which are historically used for option pricing, specifically: Black-Scholes and Heston Model. To achieve this, the study was divided into 3 parts. First pricing models were trained on simulated option price data with linear distortion. After getting results showing that models can approximate Black-Scholes and Heston Model, the models were tested on market with non-linear distortions. The final and most important stage was testing on real market data, and their performance was evaluated using different statistical error metrics. The results contribute to the ongoing debate on whether data-driven models can enhance pricing accuracy, volatility estimation, and risk management in financial markets. 

All code used in this study is available at \href{https://github.com/GeorgMil29/OptionPricing.git}{GitHub}, including data generation, model implementation, and evaluation scripts.
	
	\section{Literature review}
	
The Black-Scholes (BS) model \cite{BSart} has long been a cornerstone in financial mathematics, providing closed-form solutions for European option pricing under assumptions such as constant volatility and frictionless markets. However, these assumptions often fail to capture the complexities of real financial markets, which exhibit features like stochastic volatility, jumps, and non-linear dependencies. To address these limitations, various extensions have been proposed, including Merton’s jump-diffusion model \cite{MERTON} and Heston’s stochastic volatility model \cite{Heston}. While these models offer improvements, they remain constrained by their parametric nature and underlying assumptions.\\

In recent years, machine learning (ML) has emerged as a promising alternative for option pricing, capable of modeling complex, non-linear relationships without stringent assumptions. Early work by \cite{Hutchinson} demonstrated the potential of neural networks in approximating option prices. Building upon this foundation, recent studies have explored various ML architectures and techniques.
The study "Machine learning for option pricing: an empirical investigation of network architectures" by \cite{Mieghem} conducted an empirical investigation into different neural network architectures for option pricing, highlighting the effectiveness of generalized highway networks in capturing option price dynamics. "Leveraging Machine Learning for High-Dimensional Option Pricing within the Uncertain Volatility Model" by \cite{Goudenge} applied machine learning techniques within the Uncertain Volatility Model framework, demonstrating enhanced accuracy in high-dimensional option pricing scenarios.
Another known article "American Option Pricing using Self-Attention GRU and Shapley Value Interpretation" by \cite{Shen} proposed a method combining self-attention mechanisms with gated recurrent units (GRU) for pricing American options, achieving improved predictive performance. Additionally there exists several studies using LSTM model and Monte Carlo simulations for option pricing.\\ 

Despite these advancements, there remains a big gap in the literature concerning comprehensive comparisons between traditional models like Black-Scholes and Heston Model to various ML approaches under both simulated market distortions and real-world data conditions. Additionally, while some studies have focused on specific option types or market conditions, a evaluation across multiple ML models and market scenarios is lacking.\\

This study aims to fill this gap by conducting a thorough comparison of Neural Networks, Random Forest, and CatBoost against the Black-Scholes and Heston model. The analysis encompasses both simulated data with linear and non-linear distortions and real market data, providing a robust assessment of each model’s performance. This approach not only evaluates pricing accuracy but also examines the models adaptability to different market conditions, offering insights into their practical applicability in financial markets.\\

\section{Methodology}
The objective of this study is to evaluate the performance of machine learning models in approximating option prices compared to the traditional models, such as  Black-Scholes and Heston model. This section outlines the theoretical foundation of these traditional methods, the data generation process, the machine learning models applied, and the performance evaluation metrics. 
\subsection{Black-Scholes model}
\noindent The Black-Scholes model is the most known formula for option pricing. But the key weakness of the Black-Scholes model is that it assumes constant volatility and risk-free interest rates, which do not hold in real-world markets.\\
The Black-Scholes model assumes that the price of the underlying asset follows a geometric Brownian motion described by the stochastic differential equation $dS_t=\mu S_t dt + \sigma S_t dW_t$
where $S_t$ is the asset price at time $t$, $\mu$ is the drift rate, $\sigma$ is the volatility, and $W_t$ is a Wiener process representing standard Brownian motion.  
By constructing a risk-free portfolio and applying Itô’s Lemma, the model derives the following partial differential equation for the price $V(S,t)$ of a European option: 
$$\cfrac{\partial V}{\partial t}+\cfrac{1}{2}\sigma^2 S^2 \cfrac{\partial^2 V}{\partial S^2}+r S\cfrac{\partial V}{\partial S} - rV=0.$$ Where $r$ is the risk-free interest rate. Solving this PDE with the terminal condition for a European call option  $V(S,T) = \max(S - K, 0)$  yields the closed-form Black-Scholes formula: $$C(s,t)= S N(d_1)- Ke^{-rT}N(d_2),$$ where $d_1=
\cfrac{\ln(\frac{S}{K})+(r+\frac{\sigma^2}{2})T}{\sigma \sqrt{T}}$ and $d_2=d_1-\sigma \sqrt{T}$. This formula was implemented in Python to compare with ML models. 
\subsection{Heston Model}
The Heston model \cite{Heston} extends the Black-Scholes framework by incorporating stochastic volatility, addressing one of the primary limitations of the Black-Scholes model. While Black-Scholes assumes constant volatility, empirical evidence demonstrates that volatility itself exhibits stochastic behavior, leading to phenomena such as volatility clustering and the volatility smile observed in option markets.

In the Heston model, the underlying asset price $S_{t}$ and its variance $v_{t}$ evolve according to the following coupled stochastic differential equations:
$$dS_{t} = \mu S_{t}dt + \sqrt{v_{t}}S_{t}dW_{t}^S$$
$$dv_{t} = \kappa(\theta- v_{t})dt + \sigma\sqrt{v_{t}}dW_{t}^v$$
where $W_{t}^S$ and $W_{t}^v$ are Wiener processes with correlation $\rho$. The parameter $\kappa$ represents the rate of mean reversion, $\theta$ is the long-term variance, and $\sigma$ is the volatility of volatility. The correlation parameter $\rho$ captures the relationship between asset returns and volatility changes, typically negative in equity markets, which accounts for the leverage effect.

For the purposes of this study, the following parameter values that are commonly cited in the literature were adopted: an initial variance $v_0 = 0.04$, mean reversion speed $\kappa = 2.0$, long-term variance $\theta = 0.04$, volatility of volatility $\sigma = 0.5$, and correlation $\rho = -0.7$. These values are supported by empirical findings in \cite{cristoff} and fall within the parameter ranges recommended by \cite{Gatheral} for equity index options.

Unlike the Black-Scholes model, the Heston model does not yield a closed-form solution in the traditional sense. However, it admits a semi-analytical solution for European options via characteristic function techniques. The price of a European call option under this framework can be expressed as:

$$C(S, v, t) = S_t P_1 - K e^{-r(T-t)} P_2,$$

\noindent where  $P_1$  and  $P_2$  are risk-neutral probabilities that are computed through numerical integration involving the model’s characteristic function. In this study, option prices were computed using the AnalyticHestonEngine in QuantLib library, which implements the Heston formula with a robust numerical method that avoids instabilities during complex integration, based on the approach of \cite{Kahl}.

By allowing the volatility to evolve stochastically and to correlate with asset returns, the Heston model is able to reproduce empirical features such as volatility smiles and skews observed in market-implied volatility surfaces. This comes at the cost of increased computational complexity, but provides a more realistic representation of the dynamics governing option prices.
\subsection{Data generation}
This study relies on two broad types of data: simulated data based on financial pricing models and real market data obtained from public sources. The simulated data was constructed in two phases using both the Black-Scholes and Heston models, while the third stage involved actual market prices to test real-world applicability.

In the first two stages, synthetic data was generated by evaluating theoretical option prices across a wide grid of economically plausible input parameters. The key variables were: stock price $S$, strike price $K$, risk-free interest rate $r$, time to maturity $T$, and volatility $\sigma$. They were discretized over the following ranges: $S \in [50, 60]$, $K \in [20, 90]$, $r \in (0.00, 0.05]$ in steps of 0.01, $T \in [0.25, 2.0]$ years in monthly increments, and $\sigma \in [0.1, 0.8]$ in steps of $0.1$. The Cartesian product of these ranges formed the full parameter grid, from which a random subset of 100,000 unique combinations was sampled to ensure computational tractability while maintaining diversity in the feature space.\\

To simulate realistic deviations from theoretical pricing, stochastic distortions were added to the option values generated by the Black-Scholes and Heston formulas. These theoretical values were computed using well-established pricing models, and the distortion terms were designed to emulate imperfections commonly observed in financial markets. In Stage 1, a linear distortion was introduced by adding zero-mean Gaussian noise drawn from the distribution  $\mathcal{N}(0, 0.15)$ to the theoretical option prices. 

In Stage 2, a non-linear distortion was applied by incorporating a sinusoidal component of the form $0.2 \cdot \sin(S)$, where $S$ is the stock price. This distortion created structured, non-parametric deviations from the theoretical price, thereby providing a testbed for assessing the ability of machine learning models to recover underlying pricing relationships in the presence of systematic pricing anomalies. In both stages, the resulting distorted values were treated as the "true option price" representing what would be observed in a market environment subject to noise and pricing frictions. These prices served as the dependent variable in all subsequent training and evaluation of the machine learning models.

The simulated option prices were generated independently using both the Black-Scholes and Heston models. In the Heston framework, prices were computed using QuantLib’s AnalyticHestonEngine, with fixed model parameters grounded in empirical literature. Specifically, the initial variance  $v_0 = 0.04$, mean reversion rate $\kappa = 2.0$, long-run variance $\theta = 0.04$, volatility of volatility $\sigma = 0.5$, and correlation $\rho = -0.7$ were adopted. These values are consistent with standard practice for simulating stochastic volatility in equity markets, as supported by studies such as \cite{Gatheral} and \cite{cristoff}. The rationale behind fixing these parameters, while varying the economic inputs $S, K, r, T, \sigma$ was to emulate a strict but realistic market regime under which machine learning models could be trained and compared. The distorted prices computed from the Heston model were treated as the observed "true market prices" while the undistorted Heston model served as the benchmark for evaluating predictive performance.

In Stage 3, real-world data was collected using the Yahoo Finance API. Option chain data for Apple Inc. (AAPL) was extracted, including bid, ask, implied volatility, strike, and expiration dates. The dataset was much smaller containing only 1,087 option contracts, because in practice, the availability of high-quality option data is constrained by factors such as limited expiration dates, sparsely traded strike prices, and the need to filter out illiquid or anomalous quotes that could distort the analysis. The midpoint between bid and ask prices was used as a proxy for the observed market price, due to its stability and ability to reflect a more accurate valuation than last-traded prices. For the risk-free rate, U.S. Treasury Par Yield Curve data was employed. Because option maturities did not align exactly with standard maturities, linear interpolation was used to compute a continuous and maturity-consistent interest rate for each option. This process ensured that pricing models based on continuous-time assumptions, such as Black-Scholes and Heston, could be fairly and consistently evaluated.

\subsection{Machine Learning Models}
This study evaluates the performance of several supervised machine learning models in the context of option pricing, with particular attention to their ability to approximate complex, non-linear relationships between financial input features and option prices. The models analyzed include feedforward neural networks (MLP), Random Forest regressors, and CatBoost. Each was selected for its theoretical suitability in modeling structured yet potentially noisy financial data. To ensure robustness, the models were trained on both simulated datasets generated from Black-Scholes and Heston pricing formulas with varying distortions and real market data. Model performance was evaluated using standard error metrics on independently sampled test sets (20\% of data). Hyperparameter tuning was conducted systematically for each model to assess their optimized predictive capacity under realistic pricing conditions. \\

\subsubsection{Neural- network}

Neural networks are particularly well-suited for capturing complex, non-linear relationships in financial data. In this study, a feedforward multi-layer perceptron (MLP) was implemented using scikit-learn’s MLPRegressor. Rather than fixing the architecture manually, the model configuration, including the number of hidden layers, neurons per layer, regularization strength, and learning rate was selected through randomized hyperparameter search. This data-driven approach allowed the network to adapt its structure to different pricing environments. All models employed the ReLU activation function and were optimized using the Adam algorithm, with training capped at 1,000 iterations to ensure convergence.

A key theoretical advantage of neural networks lies in their universal approximation property, enabling them to approximate any continuous function. This makes them particularly suitable for modeling the non-linear and high-dimensional dependencies that govern option pricing in both synthetic and real market conditions.

\subsubsection{Random Forrest}
Random Forest is a non-parametric ensemble learning method based on bootstrap aggregation (bagging) of decision trees. Each tree is trained on a randomly sampled subset of the training data (with replacement), and at each split, a random subset of features is considered. This stochasticity reduces variance and decorrelates individual trees, resulting in a more stable and robust aggregated model. In this study, scikit-learn’s RandomForestRegressor was used, and key hyperparameters including the number of trees, maximum tree depth, minimum number of samples required to split a node and minimum samples per leaf  were optimized via RandomizedSearchCV. This probabilistic approach to hyperparameter tuning allowed for efficient exploration of a large search space while mitigating the computational cost associated with grid search.

Random Forest is particularly well-suited for capturing non-linear dependencies and interactions between input variables without requiring assumptions about the functional form of the underlying relationship. Its resistance to overfitting and natural handling of noisy and high-dimensional data make it an effective model for option pricing applications, especially in settings where feature importance and interpretability are also of interest.

\subsubsection{CatBoost}
Gradient boosting methods are widely regarded for their ability to model non-linear relationships and complex feature interactions with high predictive accuracy. While XGBoost was initially considered for this study due to its established performance in financial applications, CatBoost was ultimately chosen for its enhanced stability, regularization, and superior handling of overfitting in small to medium-sized datasets. Unlike traditional boosting algorithms that may suffer from variance or require extensive tuning to generalize well, CatBoost introduces improvements such as symmetric tree structures and built-in techniques for reducing overfitting, making it especially well-suited for structured numerical data with limited sample sizes.

In this study, the CatBoostRegressor was applied to the standardized feature set and incorporated into a modeling pipeline. Hyperparameter tuning was performed using Bayesian optimization (BayesSearchCV), which constructs a probabilistic model of the performance surface and iteratively selects hyperparameter configurations that balance exploration and exploitation. This method is particularly effective in high-dimensional or computationally expensive search spaces, as it converges more quickly to near-optimal configurations than grid or random search.

\subsubsection{Model Evaluation}
All models were evaluated using standard performance metrics: Mean Squared Error (MSE), Mean Absolute Error (MAE), and the coefficient of determination ($R^2$). These metrics jointly capture both the magnitude of prediction errors and the proportion of variance explained by the model. This allows for consistent and interpretable comparisons across machine learning methods and theoretical benchmarks. Feature scaling using StandardScaler was tested in preliminary experiments but did not materially improve results, and was therefore not emphasized in the final model configurations.

\section{Results and Analysis}
The results of this study are structured into three stages, each designed to assess the performance of machine learning models under different market conditions. The first stage involves synthetic data with linear distortions, simulating basic market imperfections. The second stage introduces non-linear distortions, which reflect more complex market dynamics such as volatility clustering and price jumps. Finally, the third stage evaluates model performance using real market data, where additional factors such as transaction costs and liquidity constraints come into play. This multi-stage approach provides a comprehensive comparison of machine learning models and the traditional Black-Scholes model in diverse scenarios.

\subsection*{Stage 1: Linear Distortions}
The first stage of the analysis served as a controlled experiment to assess the ability of machine learning algorithms to approximate a known pricing function under a simplified, yet systematically distorted, market environment. Synthetic data was generated using the Black-Scholes formula for European call options. The inputs were: underlying asset price $S$, strike price $K$, risk-free interest rate $r$, time to maturity $T$, and volatility $\sigma$ that were varied across plausible ranges to generate a diverse option price surface. A normally distributed noise term was then added to the theoretical prices to simulate modest market imperfections and observational error. The resulting "observed" price was treated as the "true option" price, serving as the target for comparison. Both the Black-Scholes formula and the machine learning models were evaluated based on how accurately they predicted this distorted price from the same set of input features.
\newline

The central research question in this stage was to what extent machine learning algorithms could recover the pricing rule and approximate the "true option price". Since the underlying pricing mechanism was based directly on the Black-Scholes formula, it was not anticipated that data-driven models would outperform the theoretical benchmark. Rather, the objective was more modest: to determine whether machine learning algorithms could approximate the distorted prices closely enough to justify further study on real market data.

The benchmark Black-Scholes formula, as expected, delivered highly accurate results with a mean squared error (MSE) of $0.009495$ and an  $R^2$  of $0.999919$. It is important to note that the option prices themselves ranged from approximately $0.0974$ to $44.500$, underscoring the scale against which these errors are interpreted. A simple feedforward neural network (MLP) with two hidden layers of 50 neurons each was first tested to evaluate the model’s capacity to approximate option prices. The results were surprisingly strong, with an MSE of $0.05655$ and an $R^2$ of $0.99952$. Although this was worse than the Black-Scholes formula, it nonetheless demonstrated the neural network’s potential. The aim of this stage was not to identify the best-performing ML model, but rather to explore whether ML models could come sufficiently close to theoretical pricing under distortion. Encouraged by the initial results, the MLP was further optimized using randomized hyperparameter tuning, which significantly improved the performance, reducing both MSE and MAE and achieving an $R^2$ of $0.9999755$ - outperforming the Black-Scholes benchmark. These results were unexpected, considering that the observed prices were generated directly from the Black-Scholes formula with only linear distortion.

Random Forest regressors were also evaluated. Without additional feature engineering, their performance lagged behind both the neural networks and the Black-Scholes formula. However, after introducing financial feature transformations such as moneyness ( $S/K$ ), log-moneyness, and interaction terms like $S \cdot \sigma$ and $K \cdot r$, Random Forests improved substantially, achieving an $R^2$ of $0.99961$. This showed that the model could be enhanced to approximate the pricing rule more closely. However, in Table \ref{tab:linear_distortion 1}, only the results without additional features are stated, to allow for a consistent comparison across models trained on the same inputs. Hyperparameter tuning was conducted using RandomizedSearchCV.

CatBoost, trained using Bayesian hyperparameter optimization, outperformed all other models in this setting, attaining the lowest MSE ($0.00262$) and highest $R^2$ score ($0.99998$). This confirms that gradient boosting methods are particularly effective in capturing structured noise patterns and complex non-linear dependencies.

\begin{table}[H]
	\centering
	\caption{Performance of models under linear distortion with Black-Scholes}
	\label{tab:linear_distortion 1}
	\begin{tabular}{|l|l|l|l|}
		\hline
		Linear Distortion          & \textbf{MSE}            & \textbf{MAE}   & \textbf{$R^2$} \\ \hline
		\textbf{Black and Scholes} & 0.009495                & 0.097447       & 0.999919                      \\ \hline
		\textbf{MLP}               & {\ul 0.002891}          & {\ul 0.041892} & {\ul 0.999975}                \\ \hline
		\textbf{RandomForrest}     & 0.072055                & 0.191389       & 0.999391                      \\ \hline
		\textbf{CatBoost}          & {\ul \textbf{0.002623}} & {\ul \textbf{0.038210}} & {\ul \textbf{0.999977}}                \\ \hline
	\end{tabular}
\end{table}
These machine learning algorithms were also tested against the Heston model, which is widely used in the financial literature for option pricing. The key advantage of the Heston model lies in its ability to incorporate stochastic volatility, allowing it to better reflect empirical features such as the volatility smile or skew. Unlike Black-Scholes, which assumes constant volatility, the Heston model treats volatility as a mean-reverting stochastic process.

The methodology mirrored that of the Black-Scholes case: option prices were generated using the Heston formula and then distorted by the addition of normally distributed noise. The same features ($S,K,r, T$ and $\sigma$) were used for the ML algorithms. The additional parameters specific to the Heston model ($v_0, \kappa,\theta, \rho$) were held fixed across all data points, consistent with common practice in simulation studies.
At this stage, it is not meaningful to compare the Black-Scholes and Heston models directly, since both were subjected to the same noise process. Instead, the most relevant question was whether machine learning algorithms that trained on the same features but targeting Heston prices could approximate or even outperform the Heston model.

\begin{table}[H]
	\centering
	\caption{Performance of models under linear distortion with Heston model}
	\label{tab:linear_distortion 2}
	\begin{tabular}{|l|l|l|l|}
		\hline
		Linear Distortion      & \textbf{MSE}       & \textbf{MAE} & \textbf{R\textasciicircum{}2} \\ \hline
		\textbf{Heston Model}  & 0.022722           & 0.118326     & 0.999830                      \\ \hline
		\textbf{MLP}           & 0.029181           & 0.136517     & 0.999782                      \\ \hline
		\textbf{RandomForrest} & \textbf{0.0231458 }         & \textbf{0.126406}     & \textbf{0.9998281}                     \\ \hline
		\textbf{CatBoost}      & 0.0260312 & 0.128367     & 0.9998067                     \\ \hline
	\end{tabular}
\end{table}
The Heston model itself performed very well under linear distortion, achieving an MSE of 0.0227 and an $R^2$ of 0.99983. While this was slightly worse than the Black-Scholes results, it is consistent with the additional complexity of the Heston formulation and its sensitivity to fixed parameter choices. Importantly, none of the machine learning models were able to outperform the Heston benchmark in this setting. The best result among the ML models came from the Random Forest, which achieved an $R^2$ of 0.99983, remarkably close, but still marginally below the benchmark. ML algorithms managed to approximate the price very good and  were not expected to outperform Heston Model, when the generated "true price" was that much dependent on it. \\

These results are significant not because the machine learning models learned the pricing formula itself, but because they were able to recover the underlying pricing structure from outputs that had been systematically distorted. That data-driven models were able to outperform the Black-Scholes formula even though the original data was fundamentally based on it and closely approximate the Heston model further supports the hypothesis that machine learning techniques are robust to realistic distortions. This provides strong motivation for applying such models to data with non-linear distortion (\nameref{subsec: Stage 2}) and to real market data (\nameref{subsec: Stage 3}), where the true pricing function is unknown and theoretical assumptions often fail.

\subsection*{Stage 2: Non-Linear Distortions}
\label{subsec: Stage 2}
The second stage of the study extends the controlled simulation framework by introducing non-linear distortions to the option prices. While the data generation process remains grounded in the Black-Scholes and Heston models, a deterministic sinusoidal perturbation was applied to each theoretical price: specifically, a term of the form $0.2 \cdot \sin(S)$ was added, where $S$ is the underlying stock price. This modification introduces systematic, non-parametric distortions that more realistically mimic pricing anomalies observed in real markets, such as volatility smiles, jumps, and liquidity effects, which are not captured by linear or Gaussian noise alone.

The objective of this stage was to evaluate the extent to which machine learning models can approximate option prices under structured nonlinear distortion, and whether they can outperform the original pricing formulas when evaluated against this more complex "true price". As in Stage 1, the distorted prices were treated as the target, while all models (Black-Scholes, Heston, and ML) were evaluated based on their ability to map from the same set of features: $S, K, r, T, \sigma$.\\

Under the Black-Scholes-based distortion, the benchmark model produced an MSE of 0.0173 and an $R^2$ of 0.99985. As in Stage 1, the feedforward neural network (MLP), optimized via randomized hyperparameter search, managed to outperform the benchmark, achieving an MSE of 0.00387 and an $R^2$ of 0.99997. This again supports the model’s capacity to learn structured deviations from theoretical pricing functions. CatBoost performed best, with an MSE of 0.00270 and an $R^2$ of 0.99998, while Random Forest also showed strong results, with an $R^2$ of 0.99939 confirming its utility in capturing nonlinearities in pricing dynamics (see Table~\ref{tab:non-linear_distortion 1}).

\begin{table}[H]
	\centering
	\caption{Performance of models under non-linear distortion with Black-Scholes}
	\label{tab:non-linear_distortion 1}
	\begin{tabular}{|l|l|l|l|}
		\hline
		Non-Linear Distortion & \textbf{MSE} & \textbf{MAE} & \textbf{$R^2$} \\ \hline
		\textbf{Black-Scholes} & 0.017328 & 0.12100 & 0.99985 \\ \hline
		\textbf{MLP} & \underline{0.00387} & \underline{0.047922} &\underline{0.999967} \\ \hline
		\textbf{Random Forest} & 0.072367 & 0.194306 & 0.999389 \\ \hline
		\textbf{CatBoost} & \textbf{\underline{0.002705}} & \textbf{\underline{0.038537}} & \textbf{\underline{0.999977}} \\ \hline
	\end{tabular}
\end{table}

The results for the Heston-based distortion were particularly interesting. Although the pricing model differs fundamentally from Black-Scholes by incorporating stochastic volatility, in this controlled setup the Heston model was parameterized with fixed values commonly used in financial simulations (see Methodology). As a result, its pricing output did not significantly deviate from the Black-Scholes formula across the sampled space. Once the same nonlinear distortion was applied to both models outputs, the resulting "true prices" became structurally similar, leading to nearly identical error metrics for the two benchmarks, each producing an MSE of approximately 0.0173 and an $R^2$ of 0.99987.

However, despite the similarity in benchmark performance, the machine learning models trained on the Heston-based targets exhibited different behavior compared to those trained on Black-Scholes-based data. This discrepancy arises because, although the input features remained constant, the models were trained to learn two distinct pricing functions. Since machine learning models are highly sensitive to the specific structure of their training targets, they adapted differently to the two functions leading to variation in error metrics.

\begin{table}[H]
	\centering
	\caption{Performance of models under non-linear distortion with Heston model}
	\label{tab:non-linear_distortion 2}
	\begin{tabular}{|l|l|l|l|}
		\hline
		Non-Linear Distortion & \textbf{MSE} & \textbf{MAE} & \textbf{$R^2$} \\ \hline
		\textbf{Heston Model} & 0.017328 & 0.12100 & 0.99987 \\ \hline
		\textbf{MLP} & \textbf{\underline{0.00094}} & \textbf{\underline{0.023791}} & \textbf{\underline{0.999992}} \\ \hline
		\textbf{Random Forest} & \underline{0.00841} & \underline{0.052710} & \underline{0.999937} \\ \hline
		\textbf{CatBoost} & \underline{0.002482} & \underline{0.0344744} & \underline{0.9999815} \\ \hline
	\end{tabular}
\end{table}

These findings contrast sharply with Stage 1, where none of the machine learning models were able to outperform the Heston benchmark. The critical difference lies in the structure of the distortion: while Gaussian noise introduces random fluctuations with no exploitable pattern, the sinusoidal component in Stage 2 adds a deterministic nonlinear structure to the pricing surface, something what the traditional models fail to account for. Machine learning algorithms, particularly neural networks and CatBoost, were able to leverage this structure to achieve superior predictive accuracy.

The results reinforce the robustness of data-driven approaches in the presence of systematic pricing deviations. Even when the theoretical benchmarks converged in performance due to the nature of the distortion, machine learning models adapted to the underlying patterns and achieved consistently better fits. This strongly supports their application to real-market data, where pricing irregularities are more nuanced and not fully explainable by closed-form models.

\subsection*{Stage 3: Real Market Data}
\label{subsec: Stage 3}
In the final stage of this study, the models were tested on real-world data to evaluate their practical applicability in a market environment where pricing dynamics are affected by stochastic volatility, market microstructure noise, and liquidity frictions. Historical data for Apple Inc. (AAPL) call options was retrieved from Yahoo Finance, covering a wide range of strike prices and expiration dates. For consistency, the closing stock price on January 1, 2025, was used across the dataset, and the midpoint between bid and ask prices was taken as a proxy for the observed market price. One of the fundamental challenges in working with real market data is the question \textit{"What is the true option price and how to define it?"}. While the last traded price is commonly used, it can be misleading due to illiquidity, stale quotes, or execution noise. In this study, the midpoint between the last bid and ask prices was used instead, as it more accurately reflects the consensus market valuation and mitigates the effects of transaction-specific volatility.
\newline

To enable accurate pricing in models such as Black-Scholes, which require a continuously compounded risk-free interest rate as an input, it is essential to obtain a maturity-consistent estimate of $r$ for each option contract. In this study, the risk-free rate was derived from U.S. Treasury Par Yield Curve Rates, which serve as a standard proxy for the term structure of risk-free interest rates in financial markets. However, because the time to expiration of options rarely matches the discrete set of maturities published in the yield curve, linear interpolation was employed to estimate the appropriate yield for each option’s specific maturity. This method ensures a smooth and consistent term structure, enabling precise present value discounting across the dataset. The interpolation step is particularly important to avoid systematic pricing errors that can arise from maturity mismatches, and it ensures comparability across models such as Black-Scholes and Heston, which are both sensitive to the choice of the risk-free rate.

The data was then structured analogously to the simulated settings in previous stages, with the same set of input features: the current stock price $S$, strike price $K$, time to maturity $T$, implied volatility $\sigma$, and interpolated risk-free rate $r$. The goal was to evaluate how well machine learning models can capture the real-world pricing structure using these inputs, without requiring model-specific assumptions such as lognormal returns or constant volatility.\\

The Black-Scholes model was used as the classical benchmark. However, as expected in a real-market setting where its assumptions are often violated, the model exhibited weaker performance relative to the results observed in the earlier synthetic experiments. It achieved a Mean Squared Error (MSE) of 211.59, Mean Average Error of $8.7308$  and an  $R^2$  score of 0.9379. Although the MSE appears high when compared to the near-zero errors in Stages 1 and 2, this is largely due to the nature of the data: in synthetic settings, the option prices were generated directly from the model itself with controlled distortions. In contrast, real market data contains a variety of unobservable factors such as liquidity constraints, bid-ask spreads, and stochastic volatility. Moreover, the option price range in this dataset is much broader, from 0.005 to 220.675 which naturally inflates squared error metrics. Importantly, the Mean Absolute Error (MAE) remains moderate equal to $8.7308$, and the  $R^2$  score still reflects that the model explains a substantial portion of the variance in observed prices. Overall, the performance of the Black-Scholes model in this setting is not only expected but reasonably good, given the simplicity of its assumptions relative to the complexity of actual market dynamics.

The Heston model, which accounts for stochastic volatility, was also implemented using the QuantLib library. Standard parameter values from empirical literature were used: initial variance $v_0 = 0.04$, mean reversion rate $\kappa = 2.0$, long-run variance $\theta = 0.04$, volatility of volatility $\sigma = 0.5$, and correlation $\rho = -0.7$. The model achieved a Mean Squared Error (MSE) of $10.41$, Mean Absolute Error (MAE) of 1.7464 and an $R^2$ score of 0.9969, representing a substantial improvement over the Black-Scholes formula. This is consistent with theoretical expectations, as the Heston framework captures stochastic volatility and partial correlation between asset returns and volatility, allowing it to better reflect features such as volatility smiles and skews that are commonly observed in real markets. The results highlight the value of incorporating more realistic dynamics into option pricing, even when using fixed model parameters that are not explicitly fitted to the dataset.
\newline

Machine learning models achieved superior performance across all evaluation metrics, demonstrating their ability to adapt to the complex, nonlinear dynamics present in real market data. The feedforward neural network (MLP), optimized through randomized hyperparameter tuning, achieved an MSE of 3.19 and an $R^2$ score of 0.9991. This represents a substantial improvement over both the Black-Scholes and Heston models, indicating that the neural network was able to effectively capture intricate interactions between features such as volatility, moneyness, and time to maturity that are not explicitly modeled in classical frameworks.

The Random Forest model also delivered strong results, with an MSE of $4.92$ and an  $R^2$  of 0.9986. While it did not reach the accuracy of the neural network, it significantly outperformed both benchmark models, reinforcing its utility in capturing nonlinear relationships and localized patterns in the data through ensemble averaging.

CatBoost, trained using Bayesian hyperparameter optimization, achieved the best performance among all models and across all stages of this study. On real market data, it attained an MSE of $1.77$ and an $R^2$ of $0.9995$, confirming its robustness and high precision in modeling complex option pricing structures. Notably, CatBoost consistently delivered the lowest error metrics in both simulated (linear and nonlinear distortion) and real-world datasets, outperforming all benchmark models and other machine learning algorithms. Its gradient boosting framework with native handling of feature interactions and regularization proves especially effective in high-noise, nonparametric financial environments (see Table~\ref{tab:realmarketdata} for a summary of results).
\begin{table}[H]
	\centering
	\caption{Model performance on real market data}
	\label{tab:realmarketdata}
	\begin{tabular}{|l|l|l|l|}
		\hline
		Real Market Data     & \textbf{MSE}       & \textbf{MAE} & \textbf{$R^2$} \\ \hline
		\textbf{Black-Scholes}  & 211.586           & 8.7308     & 0.9378                    \\ \hline
		\textbf{Heston Model}  & 10.41308           & 1.74647     & 0.99694                      \\ \hline
		\textbf{MLP}           & \underline{3.1946}           & \underline{0.65530}     & \underline{0.99906}                     \\ \hline
		\textbf{RandomForrest} & \underline{4.9154}         & \underline{1.02649}  &\underline{0.99855}                     \\ \hline
		\textbf{CatBoost}      & \textbf{\underline{1.767885}} & \textbf{\underline{0.776516}}   &\textbf{\underline{0.99948}}                 \\ \hline
	\end{tabular}
\end{table}
These findings are particularly compelling because the models are evaluated on market data where the true pricing function is unknown and potentially affected by multiple unobserved factors. The strong performance of machine learning models, even without explicitly modeling volatility dynamics or market frictions highlights their flexibility and robustness. Moreover, the interpolated yield curve and consistent feature set make the comparison with theoretical models both fair and meaningful. The results provide strong empirical support for the viability of machine learning approaches in practical option pricing applications, where traditional models often fall short.

\subsection*{Granular Error Analysis on Real Market Data}

\begin{figure}[ht]
	\centering
	\includegraphics[width=0.45\textwidth]{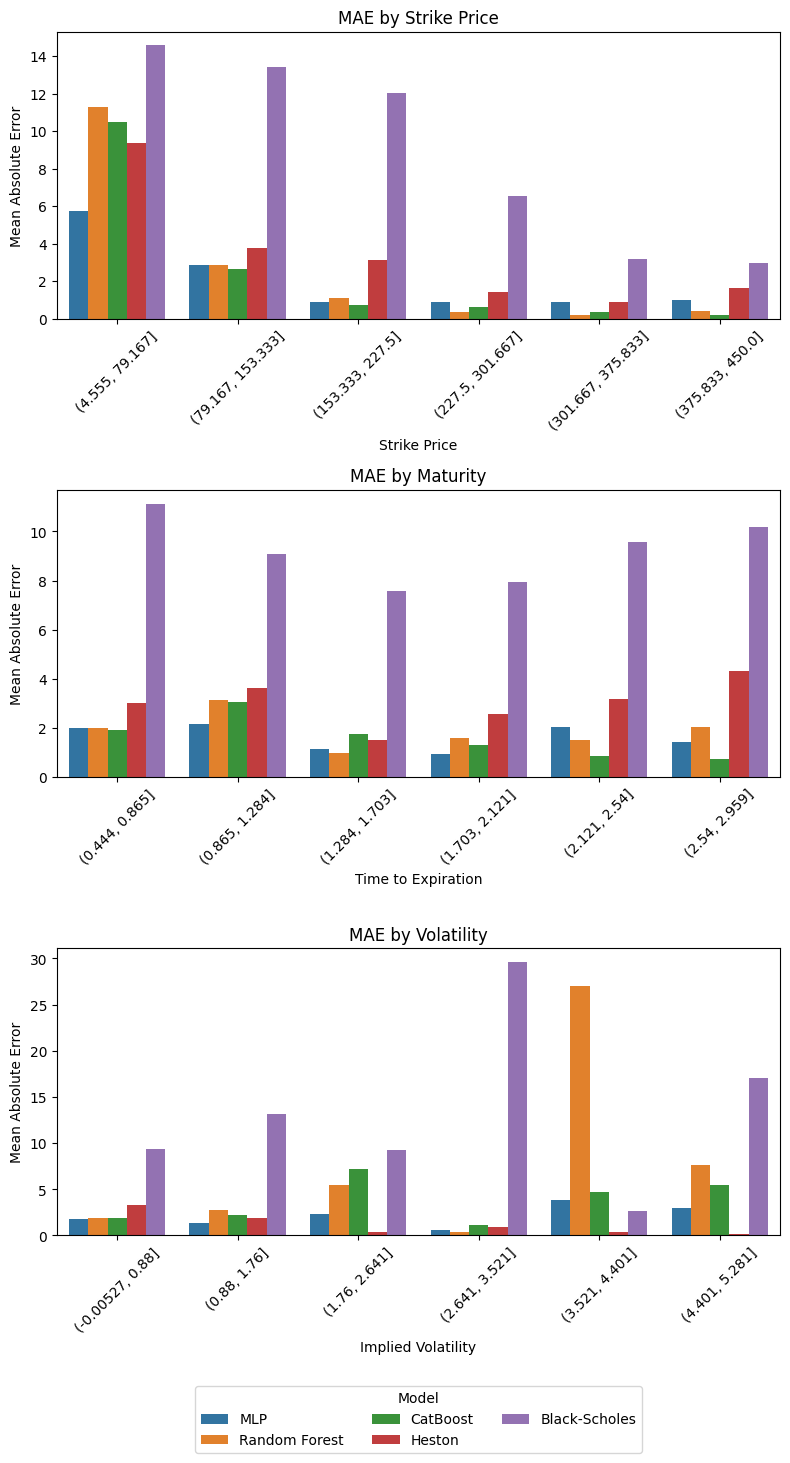}
	\caption{Mean Absolute Error (MAE) by strike price (top), time to expiration (middle), and implied volatility (bottom) for each pricing model on real market data.}
	\label{tab:bindata}
\end{figure}

While the overall error metrics in Table \ref{tab:realmarketdata} provide a clear summary of model performance, a more granular analysis is necessary to understand where these models excel or fail. To investigate this, the Mean Absolute Error (MAE) was calculated for different subsets of the real market data, binned by strike price, time to expiration, and implied volatility. Figure \ref{tab:bindata} illustrates these conditional errors, revealing the practical robustness of each model.

The analysis by strike price shows that the Black-Scholes model's performance degrades significantly for deep in-the-money options (lower strike prices), where its pricing assumptions are most strained. In contrast, while all models exhibit higher errors in this region, the machine learning models, particularly MLP and CatBoost, maintain considerably lower error levels, demonstrating their superior ability to capture the non-linear payoff structure.

The "MAE by Volatility" plot offers the most telling insight. As expected, prediction errors for all models increase with volatility. However, the performance of the Black-Scholes model deteriorates dramatically at higher volatility levels, leading to large errors. The Heston model, which is designed to handle stochastic volatility, offers a significant improvement over Black-Scholes but still shows a marked increase in error. The machine learning models, however, prove to be far more robust. CatBoost and MLP, in particular, maintain a relatively stable and low error profile even in the highest volatility regimes, underscoring their effectiveness in learning complex market dynamics without the rigid assumptions of traditional models.

\subsection*{Computational Performance on Real Market Data}
A crucial factor for the practical adoption of any pricing model is its computational cost, especially in environments requiring real-time valuation and risk management. Table \ref{tab:computional time} presents a comparison of the computational time required for both training and prediction for each model on the real market dataset.

As shown in Table \ref{tab:computional time}, the Black-Scholes model offers the fastest prediction time due to its closed-form solution, though as established, this comes at the cost of accuracy. The Heston model, requiring numerical integration to solve its characteristic function, is substantially slower in prediction.

The machine learning models introduce a necessary upfront training period, with CatBoost requiring the most time (25.6s) and Random Forest the least (3.7s). However, once trained, their prediction speeds are remarkable. The MLP and CatBoost models deliver predictions significantly faster than even the Black-Scholes model. This finding is critical: for any practical application where a model is trained periodically (e.g., overnight) and then used for repeated predictions, the ML models offer a superior combination of both high accuracy and extremely low-latency performance. This makes them highly suitable for demanding applications such as large-scale portfolio valuation and real-time hedging.
\begin{table}[H]
	\centering
	\caption{Computational time for the models}
	\label{tab:computional time}
	\begin{tabular}{|l|l|l|l|}
		\hline
		Real Market Data & \textbf{Training time (s)} & \textbf{Prediction Time (s)} \\ \hline
		\textbf{Black-Scholes} & N/A & 0.0112  \\ \hline
		\textbf{Heston Model} & N/A & 0.4641   \\ \hline
		\textbf{MLP} & {11.3393} & {0.0060}  \\ \hline
		\textbf{Random Forest} & 3.7220 & 0.0154  \\ \hline
		\textbf{CatBoost} & {25.5921} & 0.0048  \\ \hline
	\end{tabular}
\end{table}

\section{Conclusion}
This study set out to address the central question: \textit{Can machine learning algorithms outperform traditional models for option pricing?} Based on a rigorous three-stage analysis, spanning simulated data with linear and nonlinear distortions, as well as real market data, the results provide strong evidence in favor of this proposition.\\

Machine learning models, particularly feedforward neural networks and CatBoost, demonstrated the capacity to approximate complex, non-linear relationships in option prices that traditional models like Black-Scholes and Heston often fail to capture. Even when the data was generated from the Black-Scholes and Heston formulas with added noise, machine learning models were able to learn the distorted pricing surfaces with remarkable accuracy, in some cases outperforming the very model from which the data was derived.

In the real-market setting, where assumptions of constant volatility and market efficiency break down, all tested machine learning algorithms outperformed the classical benchmarks across multiple error metrics. CatBoost, in particular, consistently achieved the lowest prediction error, suggesting a strong ability to adapt to market-implied pricing structures.

Thus, the findings of this study indicate that machine learning algorithms not only offer a viable alternative but can indeed outperform traditional option pricing models under realistic conditions. While theoretical models remain important for interpretability and analytical insight, data-driven approaches offer significant advantages in flexibility, adaptability, and empirical accuracy. Furthermore, despite requiring an initial training period, the leading machine learning models offer prediction speeds that are superior to traditional methods, making them highly viable for practical, real-time applications. This opens promising directions for future research and practical implementation in derivative pricing and risk management.

\bibliographystyle{apalike}
\bibliography{biblioteka}
\end{document}